\documentclass[10pt,twocolumn,letterpaper]{article}

\usepackage[pagenumbers]{iccv} 
\usepackage{pifont}

\newcommand{\hescape}{HESCAPE}

\newcommand{\ctranspath}{CTransPath}
\newcommand{\uni}{UNI}
\newcommand{\conch}{CONCH}
\newcommand{\gigapath}{Gigapath}
\newcommand{\optimus}{H0mini}

\newcommand{\mlp}{MLP}
\newcommand{\nicheformer}{Nicheformer}
\newcommand{\drvi}{DRVI}
\newcommand{\scfoundation}{scFoundation}

\newcommand{\datafivekpanel}{Human 5K Panel}
\newcommand{\datamultitissuepanel}{Human Multi-Tissue Panel}

\newcommand{\datafivekshort}{5K}
\newcommand{\datamultitissueshort}{Multi}
\newcommand{\dataimmunooncologyshort}{ImmOnc}
\newcommand{\datacolonshort}{Colon}
\newcommand{\databreastshort}{Breast}
\newcommand{\datalungshort}{Lung}

\definecolor{iccvblue}{rgb}{0.21,0.49,0.74}
\usepackage[pagebackref,breaklinks,colorlinks,allcolors=iccvblue]{hyperref}

\usepackage[utf8]{inputenc} 
\usepackage[T1]{fontenc}    
\usepackage{hyperref}       
\usepackage{url}            
\usepackage{booktabs}       
\usepackage{amsfonts}       
\usepackage{nicefrac}       
\usepackage{microtype}      
\usepackage{xcolor}         
\usepackage[accsupp]{axessibility}

\usepackage{multirow}
\usepackage{array}

\def\paperID{45} 
\def\confName{ICCV}
\def\confYear{2025}

\title{A Large-Scale Benchmark of Cross-Modal Learning for Histology and Gene Expression in Spatial Transcriptomics}

\author{
Rushin H. Gindra$^{1,2}$\thanks{Contributed equally}, Giovanni Palla$^{1}$\footnotemark[1], Mathias Nguyen$^{1, 2}$, Sophia J. Wagner$^{1, 2}$, \\
Manuel Tran$^{1, 2}$, Fabian J Theis$^{1, 2}$, Dieter Saur$^{2}$, Lorin Crawford$^{3}$\thanks{Corresponding authors}, Tingying Peng$^{1, 2}$\footnotemark[2] \\
$^{1}$Helmholtz Munich, Germany\\
$^{2}$Technical University Munich, Germany\\
$^{3}$Microsoft Research, USA\\
{\tt\small rushin.gindra@helmholtz-munich.de, lcrawford@microsoft.com, tingying.peng@helmholtz-munich.de}
}

\begin{document}
\maketitle
\begin{abstract}
Spatial transcriptomics enables simultaneous measurement of gene expression and tissue morphology, offering unprecedented insights into cellular organization and disease mechanisms. However, the field lacks comprehensive benchmarks for evaluating multimodal learning methods that leverage both  histology images and gene expression data. Here, we present \hescape{}, a large-scale benchmark for cross-modal contrastive pretraining in spatial transcriptomics, built on a curated pan-organ dataset spanning 6 different gene panels and 54 donors. We systematically evaluated state-of-the-art image and gene expression encoders across multiple pretraining strategies and assessed their effectiveness on two downstream tasks: gene mutation classification and gene expression prediction. Our benchmark demonstrates that gene expression encoders are the primary determinant of strong representational alignment, and that gene models pretrained on spatial transcriptomics data outperform both those trained without spatial data and simple baseline approaches. However, downstream task evaluation reveals a striking contradiction: while contrastive pretraining consistently improves gene mutation classification performance, it degrades direct gene expression prediction compared to baseline encoders trained without cross-modal objectives. We identify batch effects as a key factor that interferes with effective cross-modal alignment. Our findings highlight the critical need for batch-robust multimodal learning approaches in spatial transcriptomics. To accelerate progress in this direction, we release \hescape{}, providing standardized datasets, evaluation protocols, and benchmarking tools for the community, at \hyperlink{https://github.com/peng-lab/hescape}{https://github.com/peng-lab/hescape}
\end{abstract}

\section{Introduction}
\label{sec:intro}

\begin{figure}
  \centering
  \includegraphics[width=\linewidth]{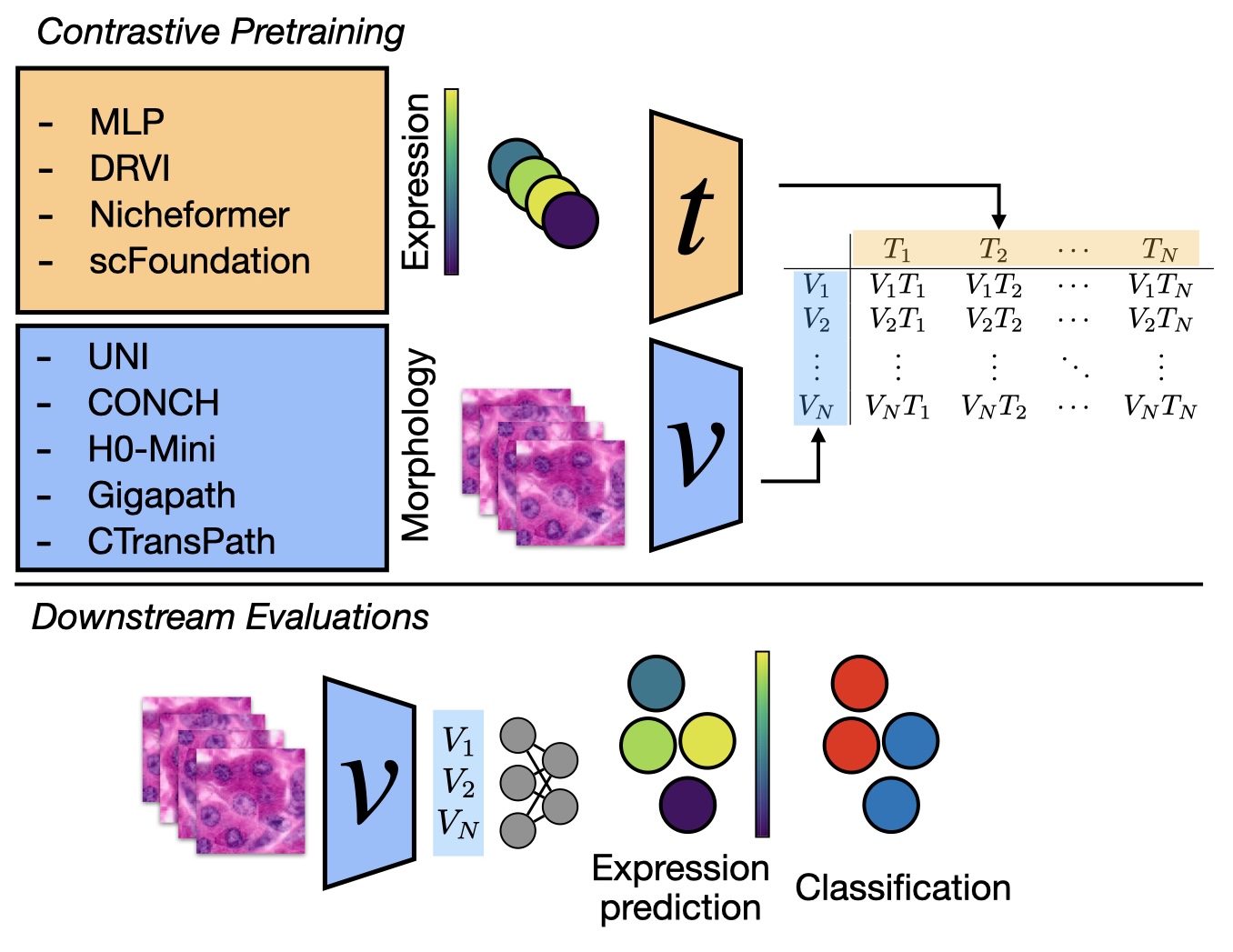}
  \caption{\hescape{} Benchmark: 4 gene expression encoders and 5 image encoders for digital pathology have been fine-tuned with contrastive pretraining, and evaluated in downstream tasks.}
  \label{fig:benchmark-schematic}
\end{figure}

Recent advances in biomedical representation learning have enabled joint modeling of pathology images with reports~\cite{lu2024conch,tran2024histogpt}, bulk gene expression~\citep{jaume2024-tangle}, and x-ray scans with clinical notes~\citep{hyland2024maira1}, largely driven by large-scale curated datasets and robust training paradigms~\citep{ding2024titan,lu2024pathchat,vaidya2025threads}.

The advent of high-throughput spatial omics techniques, including spatial transcriptomics, has opened new avenues for understanding tissue organization at cellular resolution by leveraging joint histological imaging and molecular profiling data~\cite{Rao2021-yanai-review,Palla2022-de}. Such techniques have been particularly impactful in cancer pathology research, where routine pathology analysis is paired with whole-transcriptome measurements from the same biospecimen, yielding rich multimodal molecular data with the potential of uncovering new molecular mechanisms and biomarkers~\cite{Park2023-cancer-review}. Furthermore, these data can be integrated with legacy data from pathology biobanks. 

Pathology foundation models have achieved strong performance across benchmarks~\cite{chen2024uni,vorontsov2024virchow, trident, meseguer2025benchmark, ecobenchmark, ma2025pathbench, campanella2025benchmark}. To fully harness the information provided by spatial transcriptomics, we require novel multimodal strategies capable of learning of integrating tissue images with gene expression. 

To elucidate this challenge, we curated a large-scale dataset of over $\sim620k$ patch-level image–gene pairs across multiple organs originating from 6 different 10x Xenium in-situ panels, collected across 54 patients and 57 individual samples. This dataset enable us to perform a large-scale benchmark of image and gene encoders using contrastive pretraining. Our results demonstrate that dataset-specific pretraining of gene expression encoders has the greatest impact on cross-modality retrieval performance. Additionally, we emphasize the critical importance of addressing batch effects in these encoders. For the gene expression modality specifically, the ability to directly model batch effects provides significant performance improvements in cross-modality retrieval tasks.

We directly evaluate the top-performing methods from contrastive pretraining cross-modality retrieval tasks in real-world downstream applications. Specifically, we assess how contrastive pretraining on spatial transcriptomics datasets impacts performance on two downstream tasks: mutation-specific tumor classification and gene expression prediction from morphology \cref{fig:benchmark-schematic}.

Building on these findings, we propose \hescape{} (H\&E + Spatial ContrAstive PrEtraining), a framework for contrastive pretraining on spatial transcriptomics data. \hescape{} provides a curated dataset with one-to-one mappings between histological patches and gene expression vectors, along with standardized performance metrics for evaluating new architectures. By making this open benchmark available on Hugging Face, we aim to accelerate the development of robust, clinically applicable spatial omics models that integrate seamlessly with widely adopted imaging encoders. This resource represents a step toward enabling the computer vision community to tackle domain-specific challenges in spatial transcriptomics, thereby advancing both fundamental multi-omics research and translational applications in oncology and beyond.

\section{Related work}
\label{sec:related_works}


\subsection{Image encoders}
The success of large-scale self-supervised learning combined with the availability of large numbers of unlabeled image patches from histology slides initiated the development of domain-specific feature encoders. These models are trained on millions of image patches from hundreds of thousands of patients using contrastive methods such as MoCov3 (RetCCL~\cite{wang2023retccl}, CTransPath~\cite{wang2022ctranspath}) or knowledge distillation approaches such as DINO~\cite{caron2021dino} or DINOv2~\cite{oquab2023dinov2} (Phikon, UNI~\cite{chen2024uni}, Virchow~\cite{vorontsov2024virchow}, H-Optimus-0~\cite{hoptimus0, filiot2025h0mini}, Prov-GigaPath~\cite{xu2024gigapath}, Atlas~\cite{alber2025atlasnovelpathologyfoundation}). 
These pathology-specific encoders provide better and more robust image representations, and enhance critical downstream tasks, such as slide-level biomarker prediction~\cite{wagner2023transformer} or clinical report generation~\cite{tran2024histogpt}.


\subsection{Gene encoders}
Single-cell transcriptomics consists of gene expression measurements at single-cell resolution. The data is typically sparse and contains signals reflecting both biological variation (e.g., tissue and cell state variation) and technical variation (e.g., sample preparation, sequencing depth). Variational autoencoders (VAEs) based on multilayer perceptrons (MLPs) such as scVI~\cite{lopez2018scvi} or DRVI~\cite{moinfar2024drvi} learn a low-dimensional embedding by aiming to reconstruct the gene expression at input. Recently, researchers have developed foundation models for transcriptomics (Geneformer~\cite{cui2024geneformer}, scGPT~\cite{cui2024scgpt}, scFoundation~\cite{hao2024scfoundation}, UCE~\cite{rosen2023uce}, Nicheformer~\cite{schaar2024nicheformer}, scGPT-spatial~\cite{wang2025scgpt}) that encode information from tens of millions of cells. While these models show that large-scale pretraining can provide benefit in downstream tasks in single-cell genomics, despite limited model capacity~\cite{DenAdel-scfm-dataselection}, their ability to encode molecular variation in a multimodal representation learning setting has not been tested to date.

\subsection{Multi-modal models in histopathology}
Combining pre-trained text encoders, image encoders, and gene encoders offers the opportunity to learn richer representations of multimodal patient data such as pathology reports, pathology images, and bulk RNA sequences. Recent models have proposed to align pathology images with text on patch-level (PLIP~\cite{plip}, QUILT-1M~\cite{ikezogwo2023quilt}, CONCH~\cite{lu2024conch}, MUSK~\cite{musk}) or on slide-level (PRISM~\cite{prism}, GigaPath~\cite{xu2024gigapath}, TITAN~\cite{ding2024titan}). Beyond morphological descriptions and clinical reports, pathology slides can be aligned with bulk gene expression (TANGLE~\cite{jaume2024-tangle}, FSM~\cite{fsm}, THREADS~\cite{vaidya2025threads}, SURVPATH~\cite{jaume2024-survpath}), or immunohistochemistry (MADELEINE~\cite{jaume2024-madeleine}). To take a step further, multi-modal integration with 3D spatial-omics has shown promising results (VORTEX~\cite{almagropérez2025vortex}). 

\subsection{Multi-modal in spatial-omics data}
Several models have been proposed for integrating the dual modality of spatial omics data~\cite{Palla2022-de, Rao2021-yanai-review}. A key challenge is the prediction of gene expression profiles from morphology~\cite{He2020-gexp-red,Schmauch2020-gexp-pred,Bergenstrahle2021-xfuse,Zhang2024-istar}. This task was recently proposed as a foundational benchmark for multimodal histology and gene expression integration~\cite{jaume2024hest}. Other studies suggest pretraining the multimodal representation with contrastive learning~\cite{Lee2024-pathomclip, xie2023_bleep, loki}, showing improved performance for the gene expression prediction task. Nevertheless, a large-scale evaluation of multimodal pretraining in spatial omics, which leverages different image and gene expression encoders as well as a large and diverse dataset across various transcriptomic panels, has not been reported. Additionally, the importance of performing multimodal pretraining on spatial-omics data, the correct selection of encoders and its corresponding impact on unimodal downstream task performance for histopathology is yet to be studied.

\section{Large-scale Contrastive Pretraining Benchmark For Spatial Transcriptomics}
\label{sec:benchmark}

In this work, we devised a large-scale benchmark to test the effectiveness of contrastive pretraining methods for learning multimodal representations of spatial transcriptomics data. In the following section, we describe the datasets, image encoder, and gene expression decoders employed in the benchmark.

\subsection{Dataset}
We curated a large collection of spatial transcriptomics data from 10x Genomics Xenium consisting of solely human tissues. The \hescape{} collection has 6 datasets, each identified by a subset of samples that share the same gene panels used to measure the samples. To highlight batch effects both across organs and collection sites, as well as to evaluate their effect on multi-modal learning, 3 datasets consist of organ-specific Xenium samples and 3 datasets consist of pan-organ Xenium samples. These samples have been pooled from different sites such as HEST benchmark~\cite{jaume2024hest} and other publicly available sources. It consists of a total of 729962 patches (623623 without duplicates). A full description of the dataset curation and preprocessing can be found in \cref{tab:dataset_stats} and Supplementary~\cref{supp:datasets}.

\begin{table}[t]
\centering
\caption{\hescape{} Dataset characteristics across 10x Xenium gene panels showing number of genes, patches, tissue sections, and samples/donors.}
\label{tab:dataset_stats}
\begin{tabular}{l|cccc}
\toprule
\textbf{Panel} & \textbf{Genes} & \textbf{Patches} & \textbf{Sections} & \textbf{Donors} \\
\midrule
\datafivekshort & 5001 & 178817 & 7 & 7 \\
\databreastshort & 280 & 127960 & 5 & 5 \\
\dataimmunooncologyshort & 380 & 67050 & 5 & 5 \\
\datalungshort & 343 & 56689 & 20 & 19 \\
\datacolonshort & 322 & 61067 & 5 & 3 \\
\datamultitissueshort & 377 & 132040 & 15 & 15 \\
\bottomrule
\end{tabular}
\end{table}

\subsection{Models}
We evaluate diverse combinations of image and gene encoder models using contrastive learning across all 6 datasets. To support future development, we open-source the benchmarking code and provide detailed instructions, fostering collaboration and the adoption of innovative, batch-robust gene encoder practices. The \hescape{} benchmark is designed as a plug-and-play framework, enabling seamless integration and rapid comparison of novel encoders from either modality.

\subsubsection{Image encoders}
We include five state-of-the-art image encoder models, each trained on a minimum of 15 million pathology patches collected from diverse sources. 

\textbf{\ctranspath} is a Swin Transformer trained on 15 million patches from 30,000 whole slide images (WSIs) in The Cancer Genome Atlas (TCGA) and Pathology Artificial Intelligence Platform (PAIP) public databases~\cite{wang2022ctranspath}.

\textbf{\uni} is a vision transformer (ViT) trained on 100 million patches from more than 100,000 H\&E stained WSIs spanning 20 major tissue types using the DINOv2 self-supervised framework based on knowledge distillation~\cite {chen2024uni}.

\textbf{\conch} is a ViT trained using vision-language alignment with CLIP on 1.15 million image-caption pairs curated from PubMed and medical textbooks~\cite{lu2024conch}.

\textbf{\gigapath} is a ViT model trained on approximately 1.1 billion pathology patches derived from  170,000 WSIs. Similar to \uni, its patch encoder is trained using the DINOv2 self-supervision framework~\cite{xu2024gigapath}.

\textbf{\optimus} is an 86 million parameters ViT, distilled from 1.1 billion parameters original H-Optimus-0~\cite{hoptimus0}, trained on a proprietary dataset of over 500,000 H\&E stained WSIs~\cite{filiot2025h0mini}.

\subsubsection{Gene encoders}
Despite the limited number of dedicated gene encoders for spatial technologies, we evaluate four models that are straightforward to train or fine-tune on spatial datasets. These models have either been developed as foundational models for single-cell or spatial genomics, or they have demonstrated robustness in extracting meaningful embeddings from gene expression data, whether at the single-cell or spatial level. 

\textbf{\mlp:} A multilayer perceptron (\mlp) is a standard baseline for gene expression data. Our implementation uses a simple two-layer \mlp{} with a hidden layer twice the size of the input dimension. This model is trained from scratch alongside the contrastive training on the specific gene expression dataset of interest, providing a straightforward benchmark for gene coding performance.

\textbf{\drvi:} This is an unsupervised VAE-based generative model that learns nonlinear, disentangled representations of single-cell data, designed as an extension to scVI~\cite{moinfar2024drvi}. We pretrain \drvi{} independently on each gene expression training dataset separately. 

\textbf{\scfoundation:} A transformer-based foundation model trained on a large collection of single-cell data. Since \scfoundation{} has shown its generalizable feature representations extend to other modalities like bulk RNA-seq data, we apply the \scfoundation{} model directly without additional spatial fine-tuning~\cite{hao2024scfoundation}.

\textbf{\nicheformer:} A transformer-based model based on BERT and adapted to both dissociated single-cell and targeted spatial transcriptomics data (e.g., 10x Xenium), \nicheformer{} has demonstrated exceptional zero-shot and fine-tuning performance in spatially relevant downstream tasks~\cite{schaar2024nicheformer}. We use the frozen nicheformer model in zero-shot settings for our benchmark.

\subsection{Pretraining}
Our experimental setup is based on a contrastive multi-modal pretraining framework. Each input to the vision encoder is an image patch $I_i$ surrounding a Xenium patch, and the encoder outputs a feature vector $\mathbf{v}_i \in \mathbb{R}^{d}$, where $d$ is the feature dimension. Each gene encoder processes the corresponding paired expression $G_i$, with pre-processing steps such as normalization or ranking applied when necessary (based on the encoder requirements). It yields embeddings $\mathbf{g}_i \in \mathbb{R}^{d}$, where $d$ is the same feature dimension as for the image encoders.

\subsubsection{Losses}
We leverage the foundational CLIP (Contrastive Language-Image Pre-Training)~\citep{Radford2021-clip} objective, as well as the newer SigLip (Sigmoid Loss for Language-Image Pre-training)~\cite{zhai2023-siglip} objective. 

Specifically, for a minibatch of images and gene expression pairs $\mathcal{B}=\left\{\left(I_1, G_1\right),\left(I_2, G_2\right), \ldots\right\}$, the CLIP learning objective for the image-gene expression pair \texttt{v2g} is:
\begin{equation}
    L_{\texttt{CLIP}^{\texttt{v2g}}} = -\frac{1}{|\mathcal{B}|} \sum_{i=1}^{|\mathcal{B}|} \log \frac{\exp(\tau \langle \mathbf{v}_i, \mathbf{g}_i \rangle)}{\sum_{j=1}^{|\mathcal{B}|} \exp(\tau \langle \mathbf{v}_i, \mathbf{g}_j \rangle)}
\end{equation}
where, again, $\mathbf{v}_i \in \mathbb{R}^d$ and $\mathbf{g}_i \in \mathbb{R}^d$ denote the image embeddings and gene embeddings from their respective encoders. The temperature $\tau$ is a learnable parameter.

For both losses, the global loss is
\begin{equation}
    L_{\texttt{CLIP}} = \frac{1}{2}\left[ L_{\texttt{v2g}} + L_{\texttt{g2v}} \right]
\end{equation}
where the inner product in $L_{\texttt{g2v}}$ is $\langle \mathbf{g}_i, \mathbf{v}_j \rangle$. In this work, we utilize both losses across the pretraining experiments to understand their performance in our data and batch size regime (Supplementary \cref{supptab:siglip_config}). We leverage the \textit{openclip} implementations in \citet{ilharco_gabriel_2021_openclip} for multi-GPU loss computation of CLIP and SigLip.

In the pretraining phase, we experiment with multimodal encoders both frozen and fine-tuned, where we do full fine-tuning for the \mlp~and \drvi~and parameter-efficient fine tuning (PEFT) using LoRA (Low-Rank Adaptation)\cite{hu2021loralowrankadaptationlarge} for all image models, \nicheformer, and \scfoundation. Additional description of the experimental configurations are shown in ~\cref{supp:finetuning_config}. For the final benchmark shown in \cref{tab:combined_results}, we keep all gene encoders frozen except for the \mlp{} baseline, which is trained from scratch alongside the contrastive pretraining.

\subsubsection{Metrics}
We evaluate the representation quality after image-transcriptomics alignment with Recall@$k$ for $k=\{1,5,10\}$, which measures the proportion of relevant items retrieved within the top $k$ retrieved samples. We report both image-to-gene (I2G) retrieval, where an image query retrieves its corresponding gene-expression profile, and gene-to-image (G2I) retrieval, where a gene-expression query retrieves the matching histological image.

\subsection{Downstream evaluations}

\subsubsection{Gene Mutation Prediction}
\label{subsub:gene_mut_pred}
Driver mutations are a hallmark of cancer. These genetic alterations directly promote the development and growth of cancer cells. They are often associated with tissue morphology (that is, the histological phenotype) and can be identified using deep learning models~\cite{wagner2023transformer}. In cancer diagnostics, this could eliminate costly and time-consuming molecular tests or genetic assays. For instance, in colorectal cancer (CRC), a mutation in \textit{BRAF} is associated with a poorer prognosis. A vision model that can reliably detect \textit{BRAF} mutations in routine histology images could identify high-risk patients earlier. This would guide the selection of targeted therapies and ultimately improve treatment outcomes. Here, we analyze whether pathology foundation models trained on image data alone are sufficient for biomarker detection and if they can be further improved by aligning gene expression data.

We benchmark multimodal alignment on nine biomarker detection tasks, three biomarkers for CRC: microsatellite instability (MSI), \textit{BRAF}, and \textit{KRAS} status; three biomarkers for breast adenocarcinoma (BRCA): ER, PR, and HER2 status; and three biomarkers for lung adenocarcinoma (LUAD): \textit{EGFR}, \textit{KRAS}, and \textit{TP53} status. All tasks are tested on H\&E-stained FFPE WSIs from the respective cohorts of TCGA~\cite{weinstein2013tcga,liu2018tcga}.

\subsubsection{HEST Benchmark}
Fine-tuned image encoders are evaluated on the HEST benchmark~\cite{jaume2024hest} by regressing patch-level gene expression from image patches. The benchmark comprises nine tasks that span various tissue types and cancer subtypes from both 10x xenium and visium datasets. Each task predicts the expression of the top 50 most variable genes from 224$\times$224 pixel histology patches 112 $\mu$m $\times$ 112 $\mu$m at 20x magnification. The features extracted at the patch level are then passed to a ridge regression model for gene expression prediction. 

To ensure a fair comparison and avoid data leakage, we maintained the identical train/test splits used in the pretraining phase, which vary slightly from the HEST benchmark. Further descriptions on dataset creation and train/test splits have been reported in the Supplementary \cref{supp:datasets}.

Here, we assess whether \hescape{}-aligned image encoders trained with spatial transcriptomics yield better expression prediction performance. 
We report Pearson's correlation coefficient (PCC) and mean squared error (MSE) between predicted and measured gene expression, averaged across genes and cross-validation folds (patient-level splits). 


\section{Results}
\label{sec:results}
\subsection{Contrastive Pretraining}
\label{subsec:pretraining}
Our large-scale pretraining benchmark demonstrates that the gene encoder \drvi{} emerges as the key determinant of performance improvement across all tissue panels and both image-to-gene (I2G) and gene-to-image (G2I) metrics (\cref{tab:combined_results} and Supplementary~\cref{tab:recall_1}-\cref{tab:recall_10}). Notably, \drvi{} paired with the image encoders \gigapath, \optimus, and \uni{} consistently achieves the top-performing Recall@5 scores across datasets. This superior performance is likely attributable to \drvi's pretraining on the gene expression modality of the training set and its capability to handle batch effects effectively.

When examining pre-trained gene foundation models, our findings reveal that despite their pretraining on large-scale datasets, they cannot surpass VAE-based models like \drvi{} pretrained on the dataset of interest. However, despite the training of MLP baselines on the dataset of interest, foundation models do demonstrate substantial improvements over them, with performance gains ranging from approximately 2-5x across different tissue types. Among the foundation models in our benchmark, \nicheformer{} consistently outperforms \scfoundation{} across all evaluated datasets, highlighting the advantage of pretraining on spatial transcriptomics data compared to models trained exclusively on single-cell genomics data.

The results also reveal significant heterogeneity in performance across tissue types, with lung tissue showing the highest overall performance (Recall@5 up to 0.709 for \gigapath-\drvi{} G2I) and the 5K dataset showing the most challenging retrieval task. Interestingly, while the I2G and G2I tasks generally show comparable performance patterns, some tissue-specific variations emerge, particularly in breast and lung datasets, where certain models show slightly asymmetric performance between the two retrieval directions.

Additionally, we perform several ablation studies to observe contributing factors for optimal pretraining. From these ablations, we observe, an optimal combination of \mlp{} projection, CLIP loss with fine-tunable image and gene encoders performs well across all spatial datasets. However, limited computational resources require us to perform locked gene-image tuning to improve the image encoders. The ablations have been reported in \cref{supp:ablation_study} and Supplementary \cref{supptab:projection_config}, \cref{supptab:siglip_config}, and \cref{supptab:finetuning_config}. 

\begin{table*}[t]
\centering
\caption{Complete Test Recall@5 Results for both Image-to-Gene (I2G) and Gene-to-Image (G2I) tasks across different tissue panels. Experiments with "—" indicate out of memory issues during training. Best results are in \textbf{bold}, second-best are \underline{underlined}.}
\label{tab:combined_results}
\resizebox{\textwidth}{!}{%
\begin{tabular}{l|cc|cc|cc|cc|cc|cc}
\toprule
& \multicolumn{2}{c|}{\textbf{\datafivekshort}} & \multicolumn{2}{c|}{\textbf{\datamultitissueshort}} & \multicolumn{2}{c|}{\textbf{\dataimmunooncologyshort}} & \multicolumn{2}{c|}{\textbf{\datacolonshort}} & \multicolumn{2}{c|}{\textbf{\databreastshort}} & \multicolumn{2}{c}{\textbf{\datalungshort}} \\
\textbf{Model} & I2G & G2I & I2G & G2I & I2G & G2I & I2G & G2I & I2G & G2I & I2G & G2I \\
\midrule
\mlp-\ctranspath & 0.103 & 0.106 & 0.138 & 0.098 & 0.110 & 0.094 & 0.098 & 0.122 & 0.116 & 0.117 & 0.103 & 0.126 \\
\mlp-\conch & 0.228 & 0.228 & 0.241 & 0.178 & 0.187 & 0.130 & 0.300 & 0.258 & 0.383 & 0.260 & 0.443 & 0.418 \\
\mlp-\gigapath & 0.257 & 0.257 & 0.297 & 0.215 & 0.179 & 0.132 & 0.313 & 0.297 & 0.390 & 0.288 & 0.510 & 0.493 \\
\mlp-\optimus & 0.235 & 0.235 & 0.209 & 0.153 & 0.173 & 0.119 & 0.296 & 0.291 & 0.309 & 0.235 & 0.358 & 0.336 \\
\mlp-\uni & 0.247 & 0.246 & 0.255 & 0.171 & 0.243 & 0.130 & 0.320 & 0.317 & 0.346 & 0.248 & 0.493 & 0.493 \\

\midrule
\scfoundation-\ctranspath & — & — & — & — & 0.119 & 0.126 & 0.105 & 0.098 & 0.138 & 0.122 & 0.125 & 0.121 \\
\scfoundation-\conch & — & — & — & — & 0.219 & 0.177 & 0.287 & 0.262 & 0.331 & 0.309 & 0.503 & 0.458 \\
\scfoundation-\gigapath & — & — & — & — & 0.251 & 0.207 & 0.294 & 0.249 & 0.348 & 0.365 & 0.590 & 0.543 \\
\scfoundation-\optimus & — & — & — & — & 0.206 & 0.171 & 0.315 & 0.272 & 0.388 & 0.377 & 0.427 & 0.345 \\
\scfoundation-\uni & — & — & — & — & 0.236 & 0.173 & 0.297 & 0.244 & 0.358 & 0.351 & 0.543 & 0.479 \\

\midrule
\nicheformer-\ctranspath & 0.105 & 0.115 & 0.126 & 0.117 & 0.127 & 0.127 & 0.092 & 0.106 & 0.110 & 0.126 & 0.122 & 0.138 \\
\nicheformer-\conch & 0.237 & 0.274 & 0.258 & 0.286 & 0.224 & 0.247 & 0.267 & 0.253 & 0.366 & 0.410 & 0.412 & 0.496 \\
\nicheformer-\gigapath & 0.241 & 0.255 & 0.274 & 0.285 & 0.247 & 0.267 & 0.261 & 0.269 & 0.414 & 0.447 & 0.473 & 0.554 \\
\nicheformer-\optimus & 0.243 & 0.273 & 0.261 & 0.277 & 0.212 & 0.215 & 0.290 & 0.278 & 0.418 & 0.451 & 0.424 & 0.498 \\
\nicheformer-\uni & 0.259 & 0.291 & 0.269 & 0.288 & 0.243 & 0.261 & 0.252 & 0.250 & 0.416 & 0.442 & 0.449 & 0.538 \\

\midrule
\drvi-\ctranspath & 0.106 & 0.116 & 0.116 & 0.147 & 0.138 & 0.135 & 0.111 & 0.123 & 0.162 & 0.144 & 0.143 & 0.163 \\
\drvi-\conch & 0.266 & 0.316 & 0.298 & 0.363 & 0.300 & 0.290 & 0.356 & 0.362 & 0.397 & 0.397 & 0.539 & 0.597 \\
\drvi-\gigapath & \underline{0.315} & \textbf{0.359} & \textbf{0.322} & \textbf{0.417} & \textbf{0.344} & \textbf{0.334} & 0.388 & 0.394 & \underline{0.461} & \underline{0.436} & \textbf{0.649} & \textbf{0.709} \\
\drvi-\optimus & 0.299 & 0.321 & 0.271 & 0.342 & 0.287 & 0.267 & \textbf{0.412} & \underline{0.397} & \textbf{0.465} & \textbf{0.461} & 0.562 & 0.612 \\
\drvi-\uni & \textbf{0.322} & \underline{0.341} & \underline{0.312} & \underline{0.396} & \underline{0.326} & \underline{0.318} & \underline{0.404} & \textbf{0.401} & 0.450 & 0.436 & \underline{0.610} & \underline{0.678} \\

\bottomrule
\end{tabular}%
}
\end{table*}

\subsection{Gene Mutation Prediction}
\label{subsec:gen_mut}
We tested whether contrastive pretraining with gene expression profiles provides improvements on whole-slide image (WSI) level classification tasks. To test how gene-panel specific pretraining affects the downstream performance across various organs, we choose our top 2 performing models \gigapath{}-\drvi{} and \uni{}-\drvi{} trained on the \datamultitissuepanel{}. Additionally, to show how gene-encoder selection can affect downstream tasks, for the \datafivekpanel{}, we selected \gigapath{}-\drvi{} as our top model and \uni{}-\nicheformer{} as our sub-optimal model. The \datafivekpanel{} and the \datamultitissuepanel{} were used as the common dataset for training to show that performance gains in gene mutation prediction task are not specific to certain organ datasets. 
We conducted downstream evaluations on mutation prediction for three biomarkers for CRC, three for BRCA, and three for LUAD (\cref{subsub:gene_mut_pred}). 

The results in \cref{tab:crc_results} reveal a nuanced pattern regarding the benefits of contrastive pretraining with gene expression encoders. For MSI prediction, \gigapath{}-\drvi{} achieves the best performance in F1 score (0.558), outperforming the baseline \gigapath{} model (0.484), representing a 15.3\% improvement. Similarly, for \textit{BRAF} prediction, \uni{}-\drvi{} (0.395) outperforms the respective baseline model (0.325) by 21.5\% and the best baseline model \gigapath{} (0.381) by 2.6\%.
However, for \textit{KRAS} prediction in CRC, both baseline models, \uni{} (0.495) and \gigapath{} (0.491), outperform their gene expression-pretrained counterparts, \gigapath{}-\drvi{} (0.474) and \uni{}-\drvi{} (0.364). 

\begin{table}[t]
\centering
\caption{Performance comparison of different encoders on CRC classification tasks. Results show F1 scores (mean(std)) for MSI, \textit{BRAF}, and \textit{KRAS} prediction tasks. Best results are in \textbf{bold}, second-best are \underline{underlined}.}
\label{tab:crc_results}
\resizebox{0.5\textwidth}{!}{%
\begin{tabular}{l|c|c|c}
\toprule
\textbf{Encoder} & \textbf{MSI} & \textbf{\textit{BRAF}} & \textbf{\textit{KRAS}} \\
\midrule
\gigapath & \underline{0.484(0.059)} & 0.381(0.108) & \underline{0.491(0.106)} \\
\uni & 0.472(0.099) & 0.325(0.137) & \textbf{0.495(0.088)} \\
\midrule
\textbf{\datamultitissuepanel{}} & & & \\
\midrule
\gigapath-\drvi & \textbf{0.558(0.091)} & \underline{0.394(0.104)} & 0.474(0.097) \\
\uni-\drvi & 0.465(0.098) & \textbf{0.395(0.076)} & 0.364(0.210) \\
\midrule
\textbf{\datafivekpanel{}} & & & \\
\midrule
\gigapath-\drvi & 0.473(0.093) & 0.353(0.041) & 0.421(0.098) \\
\uni-\nicheformer & 0.416(0.032) & 0.338(0.092) & 0.419(0.156) \\
\bottomrule
\end{tabular}%
}
\end{table}

Similarly, biomarker prediction in BRCA demonstrates a comparable pattern of  performance (\cref{tab:brca_results}). For ER status prediction, \gigapath{}-\drvi{} achieves the highest F1 score of 0.898, representing a slight improvement over the baseline \gigapath{} (0.890), consistent with the UNI variants. For PR status prediction, \gigapath{}-\drvi{} again shows the best performance with an F1 score of 0.829, outperforming both baseline models, \gigapath{} (0.811) and \uni{} (0.807).
However, for HER2 status prediction, the baseline models demonstrate superior performance, with \uni{} achieving the highest F1 score of 0.439, compared to the gene expression-pretrained models, \gigapath{}-\drvi{} (0.401) and \uni{}-\drvi{} (0.368). 

\begin{table}[t]
\centering
\caption{Performance comparison of different encoders on BRCA classification tasks. Results show F1 scores (mean(std)) for ER, PR, and HER2 prediction tasks. Best results are in \textbf{bold}, second-best are \underline{underlined}.}
\label{tab:brca_results}
\resizebox{0.5\textwidth}{!}{%
\begin{tabular}{l|c|c|c}
\toprule
\textbf{Encoder} & \textbf{ER} & \textbf{PR} & \textbf{HER2} \\
\midrule
\gigapath & \underline{0.890(0.022)} & 0.811(0.019) & 0.388(0.053) \\
\uni & 0.857(0.017) & 0.807(0.038) & \textbf{0.439(0.021)} \\
\midrule
\textbf{\datamultitissuepanel{}} & & & \\
\midrule
\gigapath-\drvi & \textbf{0.898(0.028)} & \textbf{0.829(0.039)} & \underline{0.401(0.042)} \\
\uni-\drvi & 0.887(0.043) & \underline{0.817(0.022)} & 0.368(0.058) \\
\midrule
\textbf{\datafivekpanel{}} & & & \\
\midrule
\gigapath-\drvi & 0.872(0.018) & 0.799(0.026) & 0.407(0.023) \\
\uni-\nicheformer & 0.858(0.022) & 0.808(0.030) & 0.433(0.023) \\
\bottomrule
\end{tabular}%
}
\end{table}

Finally, WSI gene mutation prediction in LUAD results in the same biomarker-specific pattern, though with more pronounced improvements depending on the vision backbone (\cref{tab:luad_results}). For two out of three biomarkers, pretraining with gene expression encoders demonstrates substantial performance gains. Most notably, for \textit{EGFR} prediction, \gigapath{}-\drvi{} achieves an F1 score of 0.299, representing a dramatic 100\% improvement over the baseline \gigapath{} (0.149). Similarly, for \textit{KRAS} prediction, \gigapath{}-\drvi{} from the \datafivekshort{} panel achieves the best performance with an F1 score of 0.433, outperforming the baseline \gigapath{} (0.367) by 18\%.
However, for \textit{TP53} prediction, there isn't a substantial improvement of the trained models over the baseline \gigapath{} which has an F1 score of 0.705, while both gene expression-pretrained variants score similarly with high variance across different training folds, \gigapath{}-\drvi{} (0.695) and \uni{}-\drvi{} (0.649). This pattern could be observed due to variations in self-supervised training data used for \uni{} and \gigapath{} baselines. The pattern also reinforces the biomarker-specific nature of the benefits observed across all three cancer types.

It should be noted that in the case of LUAD WSI gene mutation prediction, we only had one sample of LUAD measured with Xenium technologies for each of the two gene panels under consideration (\datafivekpanel{} and \datamultitissuepanel). Hence, this represents pretraining in a pan-cancer setting rather than lung-specific pretraining, which may explain some of the performance variations observed compared to the more tissue-specific pretraining scenarios in CRC and BRCA analyses.

\begin{table}[t]
\centering
\caption{Performance comparison of different encoders on LUAD classification tasks. Results show F1 scores (mean(std)) for \textit{EGFR}, \textit{KRAS}, and \textit{TP53} prediction tasks. Best results are in \textbf{bold}, second-best are \underline{underlined}.}
\label{tab:luad_results}
\resizebox{0.5\textwidth}{!}{%
\begin{tabular}{l|c|c|c}
\toprule
\textbf{Encoder} & \textbf{\textit{EGFR}} & \textbf{\textit{KRAS}} & \textbf{\textit{TP53}} \\
\midrule
\gigapath & 0.149(0.0797) & 0.367(0.1437) & \textbf{0.705(0.0567)} \\
\uni & \textbf{0.349(0.070)} & 0.382(0.095) & 0.693(0.057) \\
\midrule
\textbf{\datamultitissuepanel{}} & & & \\
\midrule
\gigapath-\drvi & 0.299(0.0432) & 0.371(0.0970) & 0.668(0.0662) \\
\uni-\drvi & \underline{0.329(0.0610)} & 0.379(0.0839) & 0.649(0.0769) \\
\midrule
\textbf{\datafivekpanel{}} & & & \\
\midrule
\gigapath-\drvi & 0.328(0.0383) & \textbf{0.433(0.0791)} & \underline{0.695(0.0656)} \\
\uni-\nicheformer & 0.301(0.0437) & \underline{0.387(0.1156)} & 0.658(0.0581) \\
\bottomrule
\end{tabular}%
}
\end{table}

An important limitation of this analysis is that the expression profiles of the target genes associated with the genomic abnormalities were not available in our Xenium datasets used for pretraining. For instance, when predicting \textit{TP53} mutations, the actual \textit{TP53} gene expression values were not measured in the spatial transcriptomics samples. Consequently, the models could only leverage latent correlation structures between the measured genes in the spatial transcriptomics data and the genomic abnormalities of interest, rather than direct gene-to-mutation associations.

This constraint may explain some of the observed variability in predictive performance across different biomarkers. Mutations in genes with stronger co-expression networks or clearer downstream effects on the measured gene panel are more likely to be predicted accurately through correlated expression patterns. In contrast, mutations in genes with weaker or more complex regulatory relationships to the measured genes might be harder to detect through proxy gene expression patterns. For instance, in BRCA, ESR1 gene is tightly linked to the ER receptor status, PGR to PR status and ERBB2 to HER2 receptor status. All three of these genes are included in the Multi-tissue and Prime 5K gene panels.

In summary, this analysis demonstrates that while integrating molecular knowledge into vision models can improve WSI mutation prediction for certain biomarkers, the effectiveness is highly biomarker-specific. The success appears to depend on both the strength of underlying biological relationships between histological features and gene expression variation, as well as the presence of measurable proxy signals within the available gene expression data that correlate with the target genomic abnormalities.

\subsection{Gene Expression Prediction}
\label{subsec:gexp_pred}
Next, we investigated whether contrastive pretraining improves the downstream task of predicting gene expression directly from histological images. Based on the superior retrieval performance demonstrated in \cref{tab:combined_results}, we selected the top-performing pretrained multimodal encoders and evaluated their image encoders on gene expression prediction using a ridge regression head as performed in the HEST gene expression prediction benchmark \cite{jaume2024hest}. To ensure a fair comparison and avoid data leakage, we maintained the identical train/test splits used in the pretraining phase. The gene expression prediction performance, evaluated using Pearson correlation coefficient (PCC) and mean squared error (MSE), is reported in~\cref{tab:gexp_prediction}.

\begin{table*}[t]
\centering
\caption{Gene Expression Prediction Task as given by HEST benchmark. Best results are in \textbf{bold}.}
\label{tab:gexp_prediction}
\resizebox{\textwidth}{!}{%
\begin{tabular}{l|cc|cc|cc|cc}
\toprule
& \multicolumn{2}{c|}{\textbf{\datafivekshort}} & \multicolumn{2}{c|}{\textbf{\dataimmunooncologyshort}} & \multicolumn{2}{c|}{\textbf{\databreastshort}} & \multicolumn{2}{c|}{\textbf{\datalungshort}} \\
\textbf{Model} & PCC & MSE & PCC & MSE & PCC & MSE & PCC & MSE \\
\midrule
\optimus & 0.321(0.022) & 2.726(0.831) & \textbf{0.571(0.026)} & 6.198(0.137) & \textbf{0.639(0.050)} & 15.237(3.962) & \textbf{0.628(0.013)} & \textbf{10.466(0.109)} \\
\gigapath & \textbf{0.338(0.021)} & \textbf{2.408(0.877)} & 0.546(0.030) & 6.167(0.210) & 0.607(0.049) & 15.753(5.160) & 0.609(0.024) & 10.604(0.406) \\
\uni & 0.319(0.031) & 2.535(1.164) & 0.546(0.020) & 5.984(0.163) & 0.605(0.023) & 14.428(5.404) & 0.607(0.019) & 10.705(0.593) \\
\midrule
\textbf{\datamultitissuepanel{}} & & & & & & & & \\
\midrule
\drvi-\optimus & 0.247(0.048) & 2.713(0.626) & 0.476(0.038) & 6.204(0.212) & 0.520(0.002) & 10.849(3.173) & 0.479(0.005) & 11.152(1.008) \\
\drvi-\gigapath & 0.277(0.020) & 2.461(0.712) & 0.479(0.028) & 6.011(0.075) & 0.560(0.026) & 10.600(3.593) & 0.549(0.010) & 10.984(1.182) \\
\drvi-\uni & 0.231(0.062) & 3.126(0.499) & 0.487(0.028) & 5.986(0.134) & 0.568(0.048) & \textbf{10.362(3.661)} & 0.548(0.007) & 10.467(0.217) \\
\midrule
\textbf{\datafivekpanel{}} & & & & & & & & \\
\midrule
\drvi-\optimus & 0.218(0.032) & 2.895(0.920) & 0.493(0.038) & \textbf{5.974(0.391)} & 0.519(0.039) & 11.260(3.737) & 0.501(0.005) & 11.339(0.757) \\
\drvi-\gigapath & 0.232(0.018) & 2.665(0.884) & 0.530(0.039) & 6.230(0.219) & 0.571(0.062) & 11.228(4.492) & 0.563(0.009) & 10.485(1.225) \\
\drvi-\uni & 0.264(0.035) & 2.661(0.671) & 0.495(0.023) & 6.280(0.269) & 0.592(0.065) & 11.904(5.408) & 0.543(0.000) & 10.845(1.146) \\
\bottomrule
\end{tabular}%
}
\end{table*}

Surprisingly, the results in ~\cref{tab:gexp_prediction} reveal a counterintuitive pattern: contrastive pretraining with gene expression encoders does not consistently improve direct gene expression prediction from histological images. Across all four datasets, the baseline image encoders (without gene expression pretraining) generally achieve superior or comparable performance compared to their pretrained counterparts. For instance, in the \datafivekshort{} dataset, the baseline \gigapath{} model achieves the highest PCC of both image encoders pre-trained on the \datamultitissueshort{} and \datafivekshort{} variants. This pattern is consistent across other datasets, where the baseline \optimus{} achieves the best performance in the \dataimmunooncologyshort{}, \databreastshort{}, and \datalungshort{} datasets. 

This negative result is counterintuitive, given that cross-modal pretraining should, in principle, improve the prediction of one modality from the other. We propose two possible explanations for this discrepancy from our initial hypothesis.

First, the strong batch effects present in the gene expression modality may skew the image encoder's learned representations, causing it to prioritize batch-specific patterns over generalizable biological features, thereby reducing performance on the test set. Second, we hypothesize that the alignment process itself may be detrimental to the image encoder's representational quality. During contrastive pretraining, the image encoder may be forced to discard rich morphological and spatial information that is crucial for gene expression prediction but not necessary for the contrastive task. 

\begin{figure}
    \centering
    \includegraphics[width=0.9\linewidth]{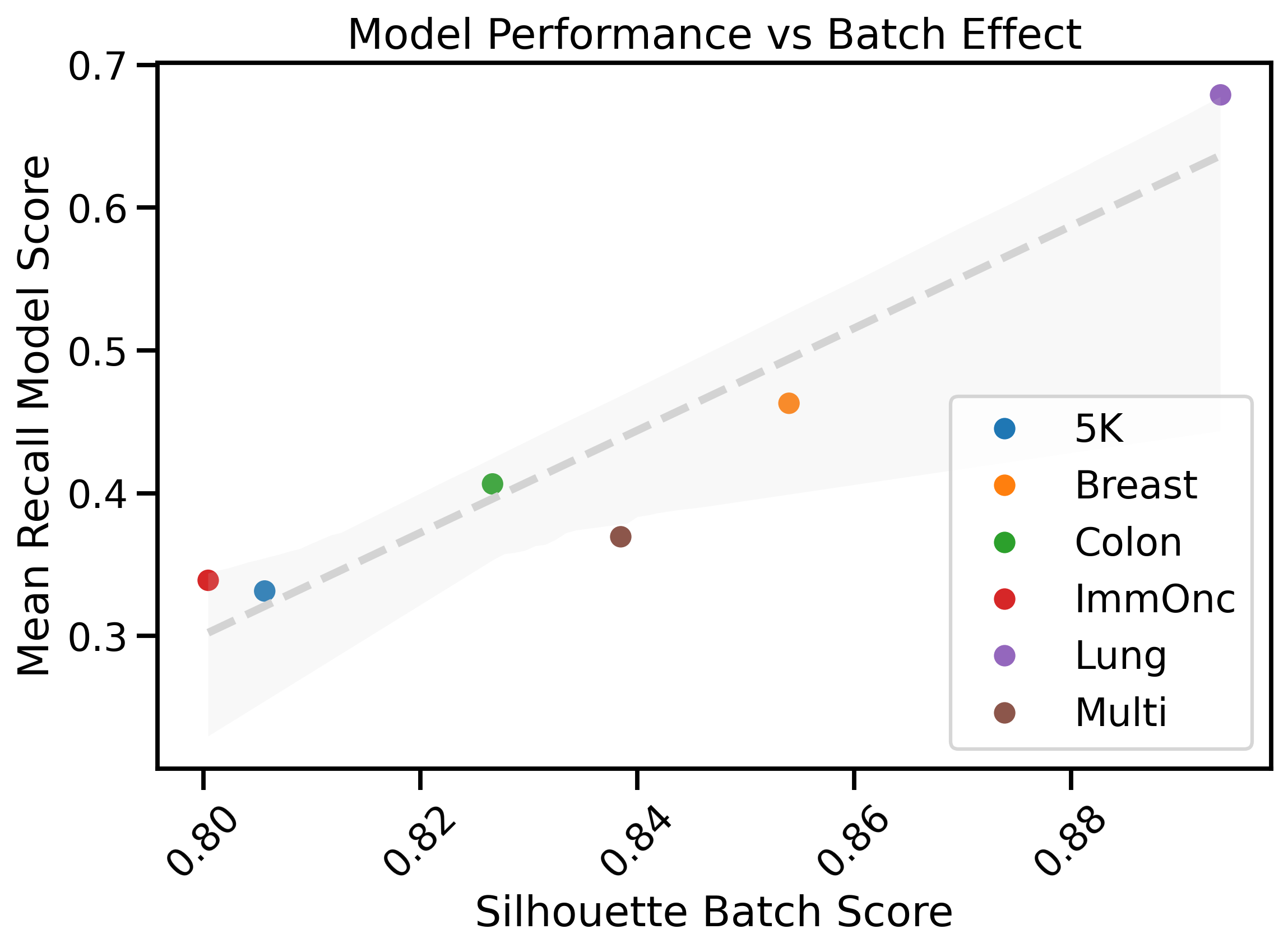}
    \caption{Silhouette Batch score against Mean Recall@5 for \gigapath-\drvi{} encoder models. Interestingly, there is a clear trend between the cross-modal retrieval task and the magnitude of batch effect in the train/val/test split.}
    \label{suppfig:batch_effect}
\end{figure}

To support this hypothesis, we present an analysis on evaluating the magnitude of batch effects in Figure~\ref{suppfig:batch_effect} and Supplementary~\cref{supp:batch_effect}, leveraging the Silhouette Batch score from \texttt{scib}~\cite{Luecken2022-integration}. 

Datasets with better batch integration (higher Silhouette Batch scores) consistently achieve superior cross-modal retrieval performance, suggesting that technical artifacts in the gene expression modality can substantially hinder the learning of meaningful cross-modal representations.

This potentially creates a trade-off between representation alignment and representation quality: while the pretrained encoder learns features that are better aligned with gene expression patterns, these aligned features may be less informative for the downstream prediction task compared to the richer, more generalizable features learned by the baseline encoder. 
In summary, the biological correlation between histological features and gene expression may be weaker than initially assumed or may be strongly affected by experimental noise such as batch effects. This creates a challenging environment for cross-modal pretraining, where strong domain-specific effects in each modality can interfere with learning meaningful cross-modal relationships. 
Our findings suggest that future work should focus on developing multimodal encoders that explicitly account for and mitigate domain-specific effects within each modality. Such approaches could potentially achieve greater robustness to batch effects while still leveraging the benefits of contrastive pretraining with spatial transcriptomics data. 
This would involve disentangling biological signal from technical noise either in an additional pretraining stage before cross-modal alignment, or online during contrastive pretraining~\cite{Cui2021-clip_limited_res}, potentially leading to more effective and generalizable multimodal representations.


\section{Conclusion}
\label{sec:conclusion}

In this work, we conducted a large-scale benchmark of contrastive pretraining encoders for cross-modal alignment of morphology and gene expression modalities in histopathology. We demonstrated improved performance on cross-modal retrieval tasks when encoders were pretrained on the dataset of interest and capable of handling batch effects appropriately. Among the foundation models tested, those directly pretrained on spatial transcriptomics data showed superior performance, highlighting the importance of including spatial transcriptomics data in the pretraining corpus for foundational single-cell genomics representations.

Furthermore, we evaluated pretrained image encoders on two downstream tasks: gene mutation classification and gene expression prediction for specific biomarkers. For mutation classification, cross-modally pretrained image encoders showed convincing improvements in performance. However, for gene expression prediction, the results were mixed, with baseline encoders often outperforming their cross-modally pretrained counterparts. This finding conflicts with findings in the literature where pretraining with gene expression appears to always improve image-to-gene expression prediction tasks~\cite{loki, xie2023_bleep, Lee2024-pathomclip}.

We hypothesize that these results are driven by strong domain-specific effects in the gene expression modality, particularly batch effects that may interfere with the learning of generalizable cross-modal representations. Our analysis provides evidence for such effects and suggests that further investigation is required to directly address domain-specific effects towards learning robust and transferable multimodal representations of morphology and gene expression. Such advances would have a significant translational impact in digital pathology and patient stratification applications.

\section{Acknowledgments}
\label{sec:acknowledgements}
R.H.G., G.P, S.J.W., M.T  were supported by the Helmholtz Association under the joint research school “Munich School for Data Science - MUDS”. S.J.W. was supported by Add-on Fellowship of the Joachim Herz Foundation. The compute resources for the project were supported by de.NBI Cloud within the German Network for Bioinformatics Infrastructure (de.NBI), ELIXIR-DE (Forschungszentrum Jülich) and Helmholtz Association's Initiative and Networking Fund on the HAICORE@FZJ partition.

{
    \small
    \bibliographystyle{ieeenat_fullname}
    \bibliography{main}
}

\clearpage

\newpage
\pagebreak

\begin{table*}[t]
\centering
\caption{Complete Test Recall@1 Results for both Image-to-Gene (I2G) and Gene-to-Image (G2I) tasks across different tissue panels. Experiments with "—" indicate out of memory issues during training}
\label{tab:recall_1}
\resizebox{\textwidth}{!}{%
\begin{tabular}{l|cc|cc|cc|cc|cc|cc}
\toprule
& \multicolumn{2}{c|}{\textbf{\datafivekshort}} & \multicolumn{2}{c|}{\textbf{\datamultitissueshort}} & \multicolumn{2}{c|}{\textbf{\dataimmunooncologyshort}} & \multicolumn{2}{c|}{\textbf{\datacolonshort}} & \multicolumn{2}{c|}{\textbf{\databreastshort}} & \multicolumn{2}{c}{\textbf{\datalungshort}} \\
\textbf{Model} & I2G & G2I & I2G & G2I & I2G & G2I & I2G & G2I & I2G & G2I & I2G & G2I \\
\midrule
\mlp-\ctranspath & 0.022(0.002) & 0.023(0.004) & 0.029(0.004) & 0.020(0.003) & 0.023(0.002) & 0.019(0.003) & 0.020(0.004) & 0.026(0.008) & 0.026(0.005) & 0.025(0.002) & 0.022(0.004) & 0.028(0.004) \\
\mlp-\conch & 0.058(0.007) & 0.058(0.006) & 0.056(0.005) & 0.041(0.005) & 0.044(0.004) & 0.028(0.002) & 0.074(0.009) & 0.062(0.005) & 0.106(0.019) & 0.067(0.011) & 0.134(0.007) & 0.127(0.013) \\
\mlp-\gigapath & 0.069(0.006) & 0.067(0.002) & 0.074(0.006) & 0.051(0.005) & 0.042(0.002) & 0.028(0.002) & 0.077(0.012) & 0.073(0.010) & 0.109(0.015) & 0.079(0.007) & 0.173(0.001) & 0.165(0.005) \\
\mlp-\optimus & 0.059(0.002) & 0.060(0.001) & 0.048(0.009) & 0.033(0.004) & 0.041(0.005) & 0.027(0.001) & 0.071(0.006) & 0.072(0.012) & 0.086(0.005) & 0.060(0.004) & 0.099(0.005) & 0.093(0.006) \\
\mlp-\uni & 0.065(0.011) & 0.064(0.009) & 0.061(0.006) & 0.039(0.006) & 0.060(0.009) & 0.029(0.006) & 0.079(0.001) & 0.080(0.007) & 0.091(0.015) & 0.065(0.014) & 0.164(0.016) & 0.164(0.019) \\

\midrule
\scfoundation-\ctranspath & — & — & — & — & 0.024(0.002) & 0.028(0.003) & 0.022(0.002) & 0.019(0.005) & 0.028(0.002) & 0.026(0.004) & 0.029(0.002) & 0.028(0.002) \\
\scfoundation-\conch & — & — & — & — & 0.055(0.006) & 0.043(0.003) & 0.077(0.008) & 0.068(0.011) & 0.085(0.012) & 0.082(0.016) & 0.166(0.005) & 0.147(0.014) \\
\scfoundation-\gigapath & — & — & — & — & 0.065(0.007) & 0.052(0.009) & 0.079(0.013) & 0.064(0.004) & 0.090(0.005) & 0.102(0.015) & 0.225(0.022) & 0.195(0.027) \\
\scfoundation-\optimus & — & — & — & — & 0.050(0.003) & 0.042(0.005) & 0.080(0.007) & 0.067(0.005) & 0.112(0.006) & 0.104(0.011) & 0.131(0.011) & 0.094(0.005) \\
\scfoundation-\uni & — & — & — & — & 0.058(0.005) & 0.040(0.004) & 0.079(0.008) & 0.060(0.021) & 0.094(0.008) & 0.094(0.015) & 0.191(0.011) & 0.158(0.011) \\

\midrule
\nicheformer-\ctranspath & 0.023(0.001) & 0.024(0.001) & 0.027(0.003) & 0.024(0.002) & 0.027(0.001) & 0.027(0.001) & 0.018(0.003) & 0.021(0.004) & 0.025(0.004) & 0.027(0.003) & 0.027(0.004) & 0.031(0.003) \\
\nicheformer-\conch & 0.061(0.001) & 0.076(0.006) & 0.067(0.005) & 0.077(0.003) & 0.056(0.003) & 0.063(0.004) & 0.069(0.006) & 0.068(0.003) & 0.101(0.008) & 0.123(0.004) & 0.125(0.002) & 0.164(0.003) \\
\nicheformer-\gigapath & 0.062(0.002) & 0.067(0.001) & 0.075(0.007) & 0.079(0.005) & 0.063(0.004) & 0.072(0.005) & 0.067(0.005) & 0.073(0.006) & 0.121(0.008) & 0.139(0.016) & 0.155(0.006) & 0.203(0.005) \\
\nicheformer-\optimus & 0.065(0.004) & 0.076(0.001) & 0.069(0.006) & 0.077(0.008) & 0.053(0.002) & 0.054(0.002) & 0.076(0.007) & 0.075(0.009) & 0.125(0.011) & 0.142(0.003) & 0.129(0.007) & 0.168(0.013) \\
\nicheformer-\uni & 0.071(0.006) & 0.082(0.003) & 0.073(0.002) & 0.079(0.002) & 0.060(0.008) & 0.067(0.007) & 0.063(0.002) & 0.066(0.004) & 0.122(0.013) & 0.136(0.004) & 0.144(0.003) & 0.188(0.006) \\

\midrule
\drvi-\ctranspath & 0.023(0.000) & 0.025(0.001) & 0.026(0.001) & 0.033(0.002) & 0.031(0.003) & 0.028(0.003) & 0.022(0.004) & 0.026(0.004) & 0.036(0.004) & 0.032(0.006) & 0.032(0.004) & 0.037(0.003) \\
\drvi-\conch & 0.073(0.005) & 0.093(0.002) & 0.084(0.000) & 0.110(0.004) & 0.083(0.005) & 0.081(0.002) & 0.098(0.010) & 0.108(0.010) & 0.117(0.022) & 0.124(0.011) & 0.195(0.004) & 0.240(0.002) \\
\drvi-\gigapath & \underline{0.097(0.011)} & \textbf{0.115(0.003)} & \textbf{0.097(0.005)} & \textbf{0.134(0.007)} & \textbf{0.103(0.004)} & \textbf{0.101(0.005)} & 0.111(0.008) & 0.121(0.010) & \underline{0.146(0.010)} & 0.142(0.008) & \textbf{0.283(0.002)} & \textbf{0.329(0.006)} \\
\drvi-\optimus & 0.085(0.005) & 0.096(0.003) & 0.077(0.008) & 0.100(0.009) & 0.079(0.006) & 0.072(0.004) & \textbf{0.123(0.001)} & \textbf{0.126(0.004)} & \textbf{0.147(0.010)} & \textbf{0.152(0.009)} & 0.208(0.001) & 0.249(0.002) \\
\drvi-\uni & \textbf{0.097(0.002)} & \underline{0.106(0.007)} & \underline{0.092(0.007)} & \underline{0.126(0.005)} & \underline{0.094(0.007)} & \underline{0.090(0.003)} & \underline{0.122(0.004)} & \underline{0.124(0.004)} & 0.142(0.023) & \underline{0.145(0.021)} & \underline{0.242(0.009)} & \underline{0.301(0.014)} \\

\bottomrule
\end{tabular}%
}
\end{table*}

\begin{table*}[t]
\centering
\caption{Complete Test Recall@10 Results for both Image-to-Gene (I2G) and Gene-to-Image (G2I) tasks across different tissue panels. Experiments with "—" indicate out of memory issues during training}
\label{tab:recall_10}
\resizebox{\textwidth}{!}{%
\begin{tabular}{l|cc|cc|cc|cc|cc|cc}
\toprule
& \multicolumn{2}{c|}{\textbf{\datafivekshort}} & \multicolumn{2}{c|}{\textbf{\datamultitissueshort}} & \multicolumn{2}{c|}{\textbf{\dataimmunooncologyshort}} & \multicolumn{2}{c|}{\textbf{\datacolonshort}} & \multicolumn{2}{c|}{\textbf{\databreastshort}} & \multicolumn{2}{c}{\textbf{\datalungshort}} \\
\textbf{Model} & I2G & G2I & I2G & G2I & I2G & G2I & I2G & G2I & I2G & G2I & I2G & G2I \\
\midrule
\mlp-\ctranspath & 0.194(0.008) & 0.203(0.017) & 0.265(0.015) & 0.191(0.011) & 0.215(0.015) & 0.183(0.008) & 0.197(0.028) & 0.235(0.081) & 0.220(0.033) & 0.220(0.016) & 0.196(0.014) & 0.234(0.039) \\
\mlp-\conch & 0.388(0.050) & 0.387(0.041) & 0.414(0.027) & 0.314(0.021) & 0.333(0.013) & 0.241(0.009) & 0.501(0.040) & 0.440(0.010) & 0.586(0.060) & 0.421(0.038) & 0.642(0.024) & 0.605(0.036) \\
\mlp-\gigapath & 0.420(0.021) & 0.425(0.015) & 0.495(0.024) & 0.371(0.019) & 0.321(0.015) & 0.244(0.008) & 0.528(0.036) & 0.500(0.028) & 0.595(0.049) & 0.441(0.036) & 0.693(0.020) & 0.675(0.016) \\
\mlp-\optimus & 0.401(0.009) & 0.399(0.008) & 0.371(0.047) & 0.279(0.029) & 0.313(0.020) & 0.224(0.013) & 0.502(0.047) & 0.492(0.058) & 0.486(0.018) & 0.384(0.017) & 0.552(0.006) & 0.523(0.003) \\
\mlp-\uni & 0.409(0.036) & 0.409(0.036) & 0.435(0.028) & 0.306(0.032) & 0.412(0.034) & 0.243(0.031) & 0.532(0.008) & 0.524(0.035) & 0.545(0.027) & 0.403(0.047) & 0.684(0.040) & 0.677(0.061) \\

\midrule
\scfoundation-\ctranspath & — & — & — & — & 0.224(0.013) & 0.232(0.017) & 0.204(0.036) & 0.192(0.037) & 0.254(0.011) & 0.236(0.020) & 0.230(0.008) & 0.225(0.017) \\
\scfoundation-\conch & — & — & — & — & 0.373(0.021) & 0.314(0.015) & 0.473(0.049) & 0.439(0.039) & 0.517(0.035) & 0.497(0.051) & 0.691(0.010) & 0.656(0.024) \\
\scfoundation-\gigapath & — & — & — & — & 0.416(0.028) & 0.352(0.029) & 0.473(0.075) & 0.422(0.014) & 0.554(0.004) & 0.567(0.030) & 0.768(0.024) & 0.733(0.043) \\
\scfoundation-\optimus & — & — & — & — & 0.358(0.013) & 0.304(0.033) & 0.513(0.004) & 0.455(0.037) & 0.581(0.021) & 0.578(0.036) & 0.616(0.019) & 0.536(0.015) \\
\scfoundation-\uni & — & — & — & — & 0.398(0.023) & 0.312(0.014) & 0.482(0.029) & 0.413(0.095) & 0.557(0.018) & 0.552(0.051) & 0.726(0.009) & 0.670(0.029) \\

\midrule
\nicheformer-\ctranspath & 0.202(0.007) & 0.218(0.005) & 0.237(0.021) & 0.224(0.030) & 0.241(0.010) & 0.233(0.011) & 0.179(0.022) & 0.212(0.026) & 0.207(0.023) & 0.242(0.018) & 0.228(0.027) & 0.256(0.030) \\
\nicheformer-\conch & 0.395(0.008) & 0.443(0.023) & 0.422(0.010) & 0.459(0.006) & 0.375(0.009) & 0.410(0.013) & 0.444(0.029) & 0.409(0.022) & 0.569(0.016) & 0.612(0.010) & 0.609(0.001) & 0.692(0.011) \\
\nicheformer-\gigapath & 0.395(0.011) & 0.417(0.011) & 0.441(0.021) & 0.458(0.017) & 0.405(0.021) & 0.434(0.006) & 0.433(0.016) & 0.439(0.038) & 0.621(0.019) & 0.648(0.027) & 0.667(0.011) & 0.742(0.006) \\
\nicheformer-\optimus & 0.396(0.014) & 0.439(0.009) & 0.430(0.023) & 0.448(0.023) & 0.355(0.009) & 0.356(0.007) & 0.470(0.038) & 0.450(0.042) & 0.620(0.018) & 0.649(0.018) & 0.617(0.013) & 0.696(0.022) \\
\nicheformer-\uni & 0.422(0.020) & 0.466(0.009) & 0.436(0.006) & 0.462(0.005) & 0.406(0.033) & 0.427(0.033) & 0.423(0.021) & 0.402(0.019) & 0.623(0.033) & 0.639(0.003) & 0.644(0.004) & 0.726(0.005) \\

\midrule
\drvi-\ctranspath & 0.204(0.006) & 0.215(0.008) & 0.222(0.013) & 0.272(0.012) & 0.258(0.015) & 0.246(0.014) & 0.212(0.025) & 0.226(0.033) & 0.289(0.015) & 0.273(0.030) & 0.264(0.017) & 0.292(0.018) \\
\drvi-\conch & 0.431(0.020) & 0.488(0.011) & 0.471(0.004) & 0.549(0.013) & 0.471(0.011) & 0.463(0.020) & 0.561(0.034) & 0.556(0.040) & 0.591(0.050) & 0.591(0.031) & 0.714(0.011) & 0.764(0.003) \\
\drvi-\gigapath & \underline{0.479(0.041)} & \textbf{0.534(0.013)} & \textbf{0.499(0.018)} & \textbf{0.607(0.015)} & \textbf{0.519(0.019)} & \textbf{0.508(0.014)} & 0.594(0.011) & \underline{0.597(0.028)} & \underline{0.660(0.024)} & 0.626(0.027) & \textbf{0.800(0.001)} & \textbf{0.849(0.003)} \\
\drvi-\optimus & 0.468(0.012) & 0.495(0.010) & 0.434(0.022) & 0.523(0.018) & 0.450(0.014) & 0.431(0.015) & \textbf{0.617(0.012)} & 0.590(0.019) & \textbf{0.664(0.023)} & \textbf{0.661(0.023)} & 0.739(0.001) & 0.778(0.004) \\
\drvi-\uni & \textbf{0.496(0.011)} & \underline{0.521(0.023)} & \underline{0.486(0.023)} & \underline{0.579(0.026)} & \underline{0.503(0.022)} & \underline{0.500(0.008)} & \underline{0.611(0.021)} & \textbf{0.600(0.013)} & 0.648(0.050) & \underline{0.629(0.045)} & \underline{0.772(0.002)} & \underline{0.826(0.008)} \\

\bottomrule
\end{tabular}%
}
\end{table*}


\section{Datasets}

This section summarizes the spatial transcriptomics datasets used in this study. The datasets are organized by gene panel type. A summary table is provided in main~\cref{tab:dataset_stats}. All tissues sections were imaged at 40x magnification and preserved using FFPE method (except where noted).


\begin{figure}
    \centering
    \includegraphics[width=0.9\linewidth]{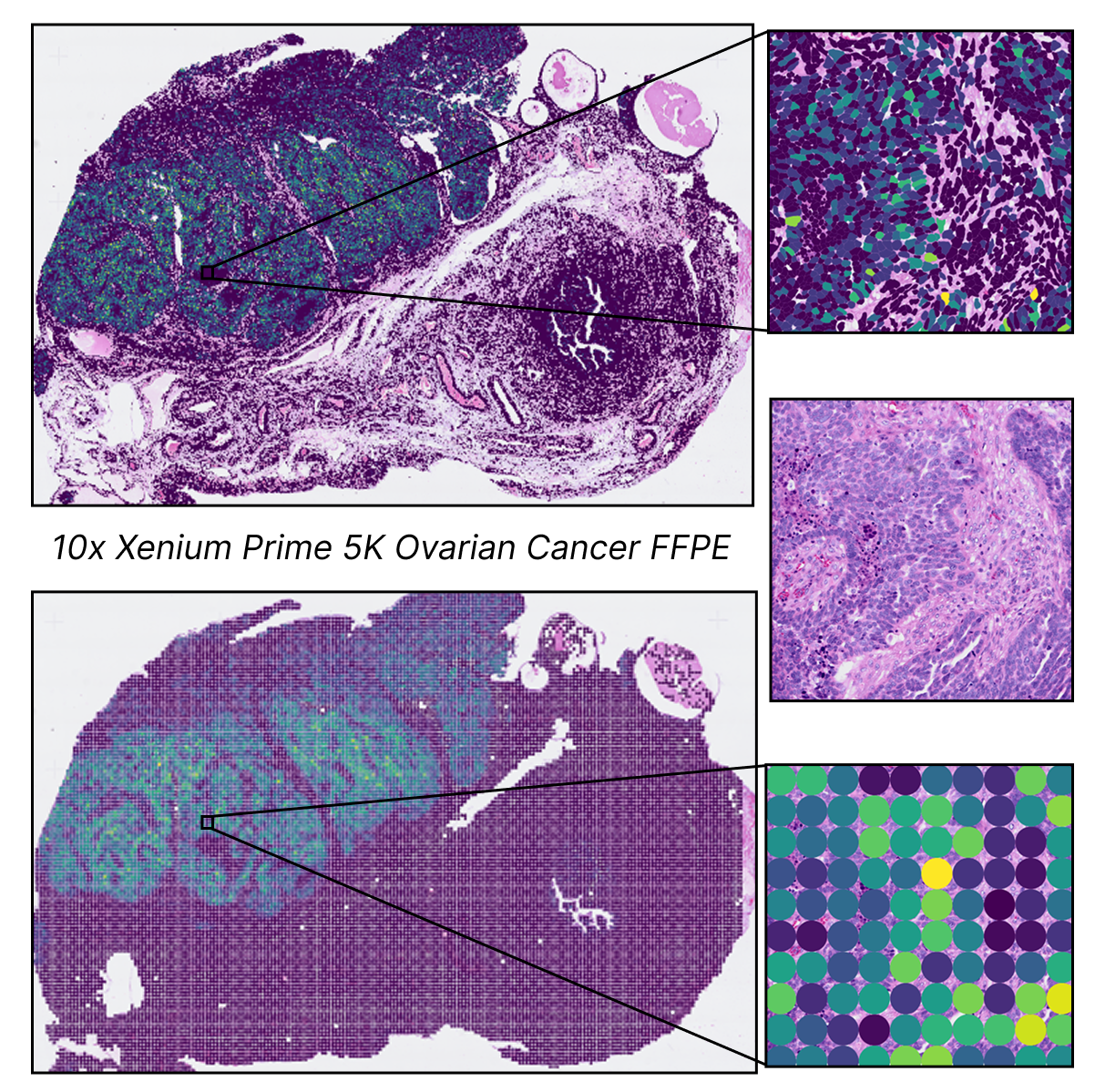}
    \caption{Top: Overlay of TP53 gene counts as observed in individual cells in Xenium Ovarian Cancer tissue. Bottom: Overlay of TP53 gene on simulated pseudo-bulk spot used to form the \hescape{} dataset}
    \label{suppfig:pseudo_bulk}
\end{figure}

Since we use 10x Xenium samples with sub-cellular transcript detection, we need to simulate a 10x Visium patch of size $55\mu m$, called pseudo-spot, for patch-based image-gene pairs. We achieve this by sum pooling the transcripts of cells within the simulated spot. The aggregated gene expressions in the simulated spots follow similar distributions to the xenium gene expressions. This is additionally validated visually by plotting expression of several biomarker genes like TP53 on the spatial samples as seen in \cref{suppfig:pseudo_bulk}.

To enable efficient benchmarking, for each dataset above, we have also created an image patch–spatial expression dataset in the Hugging Face Arrow format, comprising a total of 7229962 image-gene expression pairs combined. Each pair is accompanied by patient-specific metadata and is available on the Hugging Face Hub. For the final benchmark, we filter out replicates with different gene panels and single gene panel datasets resulting in $\sim620k$ image-gene expression pairs.

Patient-based stratification was employed for each dataset to create training, validation, and test splits-ensuring that each patient’s samples are confined to a single split. Since a lot of 10x Xenium samples are part of the HEST-benchmark test set, for fair evaluation we design our own testing splits and test all image models from HEST-benchmark on these new splits. All reported results are based on the provided test split to clearly expose the impact of batch effects.

All dataset preprocessing and creation was done using SpatialData \cite{spatialdata} and Huggingface Datasets library

\subsection{Detailed Description of Dataset Groups}
\label{supp:datasets}
\subsubsection{Human 5K Panel}
This panel contains 6 datasets spanning multiple organs including Skin, Prostate, Lymphoid, Lung, Breast, Cervix, and Ovary. All samples are from cancer or diseased tissue preserved with FFPE, except one Fresh Frozen ovarian cancer sample. The datasets have uniform pixel size around 0.274 $\mu$m. 

\subsubsection{Human Colon Panel}
This panel includes 5 datasets exclusively focused on bowel tissue, with both cancerous and healthy samples. All samples are FFPE-preserved with pixel sizes ranging from 0.137 $\mu$m to 0.274 $\mu$m. Three datasets come from the same study on immune cell populations in colorectal cancer, all from the same patient.

\subsubsection{Human Lung Healthy Panel}
The most extensive group with 19 datasets, all focused on lung tissue. Samples include both healthy (6 datasets) and diseased (13 datasets) states from 19 different patients with pulmonary fibrosis. All datasets are from the study Vannan et al.~\cite{lung-datasets-vannan}. All images have consistent pixel size of 0.213 $\mu$m. 

\subsubsection{Human Immuno-Oncology Panel}
This panel contains 5 datasets from 5 different organs (Ovary, Lung, Pancreas, Bowel, Brain), all studying cancerous tissue. Three datasets include identified patient information. All samples have consistent pixel size around 0.274 $\mu$m and include custom add-ons to the base panel.

\subsubsection{Human Multi-Tissue Panel}
A diverse group of 14 datasets covering 8 different organs (Lymphoid, Bone, Pancreas, Skin, Liver, Heart, Lung, Kidney). Includes all three disease states (Cancer, Healthy, Diseased) with pixel sizes ranging from 0.137 $\mu$m to 0.274 $\mu$m. Most datasets are organ-specific paired samples (cancer/healthy or diseased/healthy). 


\subsubsection{Human Breast Panel}
Includes 5 datasets all focused on breast cancer tissue. Two datasets are from a study using the entire sample area from one patient, while three are from a high-resolution mapping study of tumor microenvironment. Pixel sizes range from 0.213 $\mu$m to 0.364 $\mu$m. 

\subsection{Additional datasets}
We also include 4 datasets studying breast cancer tissue from two patients in the \hescape{} dataset. These datasets were not used for any experiments above, but are can be useful for further downstream tasks. Each patient has two datasets: one with a custom add-on panel and one with a pre-designed panel, allowing for direct comparison. All datasets have consistent pixel size of 0.213 $\mu$m.


\subsubsection{Preservation Methods}
Only one dataset, Xenium\_Prime\_Human\_Ovary\_FF, uses Fresh Frozen preservation for ovarian cancer tissue.

\subsubsection{Imaging Parameters}
All datasets were imaged at 40x magnification. Pixel sizes range from 0.137 $\mu$m to 0.364 $\mu$m, with most datasets having pixel sizes around 0.213 $\mu$m or 0.274 $\mu$m.

\section{Pretraining}
\label{supp:pretraining}

\subsection{Implementation details}
For consistency, both encoders output embeddings of dimension $d=128$. The training is performed with gradient clipping set to $5.0$, a batch size of $256$ distributed across $4$ GPUs for $20,000$ steps, and an initial warmup phase of 780 steps. We use an AdamW optimizer $\beta\colon(0.9, 0.95)$ with a learning rate starting at $1 \times 10^{-5}$ reduced over iterations via a cosine scheduler, and a weight decay of $0.01$. The full contrastive pretraining is performed with a finetuning image-tuning and locked gene encoding objective, allowing the image encoders to learn robust, gene-biomarker specific features. All experiments are conducted on a Slurm GPU cluster equipped with A100 GPUs. We use PyTorch and Hydra for all our experiments.

\begin{table}[t]
\centering
\caption{Projection head ablation study: Experiments performed with CLIP loss and frozen encoders. Best results are in \textbf{bold}, second-best are \underline{underlined}.}
\label{supptab:projection_config}
\resizebox{0.5\textwidth}{!}{%
\begin{tabular}{l|c|cc|cc}
\toprule
 &  & \multicolumn{2}{c|}{\textbf{\datafivekshort}} & \multicolumn{2}{c}{\textbf{\datacolonshort}} \\
\textbf{Model} & \textbf{projection} & I2G & G2I & I2G & G2I \\
\midrule
\drvi-\gigapath & linear & 0.170 & 0.180 & \underline{0.295} & 0.274 \\
\drvi-\uni & linear & 0.172 & 0.180 & 0.243 & 0.251 \\
\midrule
\drvi-\gigapath & mlp & \textbf{0.201} & \textbf{0.210} & \textbf{0.317} & \textbf{0.328} \\
\drvi-\uni & mlp & \textbf{0.201} & \underline{0.201} & 0.289 & 0.231 \\
\midrule
\drvi-\gigapath & transformer & 0.114 & 0.130 & 0.250 & \underline{0.277} \\
\drvi-\uni & transformer & 0.135 & 0.166 & 0.179 & 0.245 \\
\bottomrule
\end{tabular}%
}
\end{table}

\begin{table}[t]
\centering
\caption{Loss function ablation study: Experiments performed with same configuration as the benchmark. Best results are in \textbf{bold}, second-best are \underline{underlined}.}
\label{supptab:siglip_config}
\resizebox{0.5\textwidth}{!}{%
\begin{tabular}{l|c|cc|cc}
\toprule
 &  & \multicolumn{2}{c|}{\textbf{\datafivekshort}} & \multicolumn{2}{c}{\textbf{\datacolonshort}} \\
\textbf{Model} & \textbf{loss} & I2G & G2I & I2G & G2I \\
\midrule
\drvi-\gigapath & CLIP & \underline{0.315} & \textbf{0.359} & \underline{0.388} & \underline{0.394} \\
\drvi-\uni & CLIP & \textbf{0.322} & 0.341 & \textbf{0.404} & \textbf{0.401} \\
\midrule
\drvi-\gigapath & SIGLIP & \textbf{0.322} & \underline{0.352} & 0.377 & 0.345 \\
\drvi-\uni & SIGLIP & 0.292 & 0.293 & 0.359 & 0.359 \\

\bottomrule
\end{tabular}%
}
\end{table}

\begin{table}[t]
\centering
\caption{Encoder Finetuning ablation study: Experiments performed with CLIP loss and MLP projection head. Experiments with "—" indicate out of memory issues during training. Best results are in \textbf{bold}, second-best are \underline{underlined}.}
\label{supptab:finetuning_config}
\resizebox{0.5\textwidth}{!}{%
\begin{tabular}{l|cc|cc|cc}
\toprule

& \multicolumn{2}{c|}{\textbf{finetune}} & \multicolumn{2}{c|}{\textbf{\datafivekshort}} & \multicolumn{2}{c}{\textbf{\datacolonshort}} \\
\textbf{Model} & img & gene & I2G & G2I & I2G & G2I \\
\midrule
\nicheformer-\uni & \ding{55} & \ding{55} & 0.153 & 0.174 & 0.201 & 0.221 \\
\nicheformer-\gigapath & \ding{55} & \ding{55} & 0.149 & 0.169 & 0.195 & 0.221 \\
\drvi-\uni & \ding{55} & \ding{55} & 0.172 & 0.180 & 0.243 & 0.251 \\
\drvi-\gigapath & \ding{55} & \ding{55} & 0.170 & 0.180 & 0.295 & 0.274 \\

\nicheformer-\uni & \ding{55} & \checkmark & 0.187 & 0.205 & 0.261 & 0.266 \\
\nicheformer-\gigapath & \ding{55} & \checkmark & 0.188 & 0.197 & 0.282 & 0.292 \\
\drvi-\uni & \ding{55} & \checkmark & 0.204 & 0.225 & 0.318 & 0.319 \\
\drvi-\gigapath & \ding{55} & \checkmark & 0.198 & 0.208 & 0.311 & 0.343 \\

\nicheformer-\uni & \checkmark & \ding{55} & 0.262 & 0.282 & 0.238 & 0.244 \\
\nicheformer-\gigapath & \checkmark & \ding{55} & 0.277 & 0.296 & 0.249 & 0.266 \\
\drvi-\uni & \checkmark & \ding{55} & 0.289 & 0.293 & 0.326 & 0.284 \\
\drvi-\gigapath & \checkmark & \ding{55} & 0.269 & 0.335 & 0.333 & \underline{0.362} \\

\nicheformer-\uni & \checkmark & \checkmark & \underline{0.308} & 0.317 & 0.323 & 0.326 \\
\nicheformer-\gigapath & \checkmark & \checkmark & — & — & — & — \\
\drvi-\uni & \checkmark & \checkmark & \textbf{0.358} & \underline{0.342} & \textbf{0.335} & 0.336 \\
\drvi-\gigapath & \checkmark & \checkmark & 0.299 & \textbf{0.370} & \underline{0.334} & \textbf{0.376} \\
\bottomrule
\end{tabular}%
}
\end{table}

\subsection{Ablation study}
\label{supp:ablation_study}
To measure the contributing factors of different hyperparameters and architectural changes towards contrastive alignment performance, we performed 3 independent ablation studies. The ablations were performed on the pan-organ 5K gene panel and the Colon panel. 

\subsubsection{Image Projection head}
\label{supp:projection_config}
The ablation was to see which of the 3 heads, the basic linear projection, MLP or a transformer based projection helps improve the image encodings during contrastive pretraining. The results in \cref{supptab:projection_config} show, that MLP as an image projection head performs consistently well across both datasets.

\subsubsection{Loss function}
\label{supp:siglip_config}
Here, we test both CLIP and SigLip losses across the 5K and Colon pretraining experiments to understand their performance in our data and batch size configurations. Our ablations in \cref{supptab:siglip_config} suggests, there was no substantial improvement from using SigLip as the loss function. 

The SigLip loss for the image-gene expression pair \texttt{v2g} is:
\begin{equation}
    L_{\texttt{SIGLIP}^{\texttt{v2g}}} = -\frac{1}{|\mathcal{B}|} \sum_{i=1}^{|\mathcal{B}|} \sum_{j=1}^{|\mathcal{B}|} \log \frac{1}{1+\exp[z_{ij}(-\tau \langle \mathbf{v}_i, \mathbf{g}_j \rangle + b)]}
\end{equation}s
with an additional learnable bias $b$. Unlike CLIP loss, SIGLIP avoids computing a global normalization and instead formalizes the objective as a logistic regression task, where the label $z_{ij}$ is 1 for the positive pair and is -1 for all the other pairs.

\subsubsection{Encoder Finetuning}
\label{supp:finetuning_config}
In the contrastive pretraining stage, HESCAPE is capable of performing both full fine-tuning for small models and parameter-efficient fine tuning (PEFT) for large transformer-based models. These approaches help to align the pretrained models to the other modality, while potentially helping the encoders to adapt to specific tasks and potentially mitigating problems arising from batch effects. 

To evaluate whether frozen pretrained encoders alone are sufficient for multimodal image-gene alignment, we conducted ablation experiments using various combinations of image-gene finetuning. We fine-tune the self-attention query-key-value embeddings and projection layers of the image encoder using LoRA~\cite{hu2021loralowrankadaptationlarge}.  In our ablation study, we find that both unlocked image and gene models enable better image-gene alignment when possible \cref{supptab:finetuning_config}. However, we can often be restricted by the compute resources for large multimodal Foundation Model finetuning. Additionally, since the gene modality is deeply affected by batch effects, we decided to keep the gene models frozen for the \hescape{} benchmark.

\subsection{Batch effects on gene expression modality}
\label{supp:batch_effect}

During dataset curation and preprocessing, we observed strong batch effects across samples of the same tissue type in all datasets under consideration. To systematically investigate how these batch effects impact contrastive pretraining performance, we employed the silhouette-batch metric from the single-cell integration benchmark \texttt{scib}~\cite{Luecken2022-integration}.

We computed this metric for all datasets using Leiden clustering results after standard scanpy preprocessing~\cite{Wolf2018-scanpy} as the \textit{label key}, and the train-validation-test split as the \textit{batch key}. By treating the dataset split as batch information, we quantified how well integrated the gene expression profiles are across different data splits, a measure that reflects the presence of technical artifacts in the gene expression modality.

\cref{suppfig:batch_effect} shows the relationship between silhouette-batch values and average Recall@5 performance for the \gigapath{}-\drvi{} model across all datasets. Notably, we observe a clear linear relationship between retrieval performance and the batch integration metric, directly supporting our hypothesis that batch effects significantly impact contrastive pretraining effectiveness. Specifically, our analysis confirms that cancer tissues exhibit significant heterogeneity in cellular composition and transcriptomic profiles relative to healthy or non-cancerous diseased samples. Our findings reveal pronounced batch effects in cancer samples as seen in the organ specific datasets, especially in breast and colon tissues, whereas lung samples with homogeneous disease conditions, in particular, patients with pulmonary fibrosis, show minimal batch variation. Technical variability can further exacerbating these differences - 10x Visium samples, for instance, are particularly prone to batch effects compared to Xenium. 

\subsection{Downstream task: gene mutation prediction}

For the evaluation of gene mutation prediction, we use a weakly-supervised learning approach for predicting the slide-level mutation targets from frozen patch-level embeddings of the pretrained models. 

For feature extraction, we use the pipeline TRIDENT\footnote{\hyperlink{https://github.com/mahmoodlab/TRIDENT}{https://github.com/mahmoodlab/TRIDENT}} with default parameters, extracting embeddings from patches of size 256$\times$256 pixels at 20$\times$ magnification. The pretrained HESCAPE models have a latent dimension of 128. 

For slide-level mutation prediction, we follow the HistoBistro pipeline\footnote{\hyperlink{https://github.com/peng-lab/HistoBistro}{https://github.com/peng-lab/HistoBistro}}.
Concretely, we employ Transformer-based feature aggregation using a two-layer Transformer architecture with eight heads of dimension 64 and latent dimension 512~\cite{wagner2023transformer}. We split the TCGA cohorts into five site-preserving folds for five-fold cross validation, using three folds for training, one for validation, and one for testing. We train the models for 10 epochs using the optimizer AdamW with learning rate of $2\times 10^{-5}$, weight decay of $2\times 10^{-5}$, and batch size 1. The best model is chosen based on the validation loss, evaluated every 500 iterations.

\end{document}